\documentstyle[pra,tighten,aps]{revtex}
\begin{document}
\draft

\title{\bf Entanglement, local measurements, and
symmetry}

\author{A A Klyachko and A S Shumovsky}

\address{Faculty of Science, Bilkent University, Bilkent, Ankara,
06800, Turkey}

\maketitle

\begin{abstract}
A definition of entanglement in terms of local measurements is
discussed. Viz, the maximum entanglement corresponds to the states
that cause the highest level of quantum fluctuations in all local
measurements determined by the dynamic symmetry group of the
system. A number of examples illustrating this definition is
considered.
\end{abstract}


\section{Introduction}

Celebrating the Centenary of Eugene Paul Wigner, one cannot but
reward  Wigner's approach to quantum mechanics that has been
formulated in his famous papers (Wigner 1931, Wigner, 1939).
According to this approach, the general properties of a quantum
mechanical system are specified by the dynamical symmetry of the
corresponding Hilbert space. For years, the approach has been used
in quantum mechanics and quantum field theory and has demonstrated
an "unexpected efficiency" (Wigner 1967). The main aim of this
paper is to apply Wigner's approach to the investigation of the
phenomenon of quantum entanglement.

It has been recognized that the notion of entanglement has a deep
conceptual meaning, touching on the problems of locality and
reality in quantum mechanics. At the same time, entanglement is
considered to be a base of quantum computing, communications, and
cryptography (see: Bowmeester {\it et al} 2000 and Tombesi and
Hirota 2001 and references therein). In spite of a great success
in engineered entanglement (for a recent review, see: Zeilinger
1999, Raimond {\it et al} 2001, Gisin {\it et al} 2002), there is
still no agreement of opinion among the experts on the very
definition of entanglement and its proper measure (e.g., see:
Peres 1998, Verdal and Plenio 1998, Brukner {\it et al} 2001).

In the usual treatment, the entanglement is associated with
nonseparability of corresponding states. It should be stressed
that the nonseparability is not a sufficient condition of maximum
entanglement, and probably of the entanglement at all (Horodecki
{\it et al} 1998).

It has been shown recently (Can {\it et al} 2002 (a)) that the
entangled states of physical systems obey a certain condition, viz
the local measurements have the maximum uncertainty in comparison
with the other states allowed for a system under consideration.
This condition can be used as an operational definition of maximum
entanglement (definition in terms of what can be measured). At the
same time, there is an adequate mathematics hidden behind this
physical definition that has been unveiled recently (Klyachko
2002).

Let us note first that this novel definition has a deep physical
meaning. One can choose to interpret the entangled state shared
between Alice and Bob as a quantum communication channel, in which
the information is carried mostly by the correlations between the
sides of the channel (Brukner {\it et al} 2001). These
correlations manifest themselves in terms of local measurements
performed at the ends of the channel and the maximum correlation
corresponds to the maximum uncertainty of local measurements
(Klyachko and Shumovsky 2002).

We now note  that the set of possible independent measurements for
a given system is specified by the symmetry properties of the
Hilbert space, corresponding to this physical system. Beginning
with this fact, reflecting the key idea of Wigner's approach, it
is possible to examine the notion of entanglement in terms of the
geometric invariant theory (Klyachko 2002) (for references on
geometric invariant theory see Mumford {\it et al} 1994).

In this paper we continue the discussion of the new definition of
maximum entanglement and consider a number of physical examples.

The paper is arranged as follows. In section II, wee consider the
definition of entangled states in terms of the maximum uncertainty
of local measurements. This definition is illustrated by a number
of examples involving the two- and three-qubit systems. In section
III, we show that the above definition of entanglement can also be
expressed in terms of a certain property of the matrix of
coefficients, specifying the entangled state. Viz, the parallel
slices of this matrix should be orthogonal and should have the
same measure. In section IV, we consider a realization of
long-lived, easy monitored entanglement in a system of three-level
$\Lambda$-type atoms. Finally, in section V we briefly discuss the
obtained results and their implementation.

\section{Definition of entanglement}

In the usual treatment, the entanglement is associated with the
states of the composite systems. Consider a composite system
defined in the Hilbert space
\begin{eqnarray}
{\cal H}= \bigotimes_{\ell =1}^N {\cal H}_{\ell}, \label{1}
\end{eqnarray}
where $N \geq 2$ is the number of components and each component
has the dimension $n_{\ell}$ (the number of independent quantum
degrees of freedom). Then, the dynamic symmetry group,
corresponding to a component, is
\begin{eqnarray}
G_{\ell}=SU(n_{\ell}). \label{2}
\end{eqnarray}
An example of some considerable interest is provided by the
$N$-qubit system, consisting of the spin-$1/2$ particles. In this
case, for all $\ell$, $n_{\ell}=2$ and $G_{\ell}=SU(2)$.

The local measurements, providing the information about
entanglement, are defined by the observables $g_{\ell}$ from the
Lie algebra $Lie G_{\ell}$ of the dynamic symmetry group
$G_{\ell}$ (2).

Assume that $g_{\ell} \in Lie \quad G{\ell}$ is a local
measurement that gives the spin projection on a given axis. Then,
the result of the local measurements is specified by the
expectation values
\begin{eqnarray}
\langle g_{\ell} \rangle = \langle \psi |g_{\ell}| \psi \rangle
\label {3}
\end{eqnarray}
and by the variances
\begin{eqnarray}
\langle (\Delta g_{\ell})^2 \rangle = \langle
\psi|(g_{\ell})^2|\psi \rangle - \langle \psi|g_{\ell}| \psi
\rangle^2 , \label{4}
\end{eqnarray}
determining the quantum error of measurements. Here $| \psi
\rangle$ denotes a state in (1).

Consider the variance (4). First of all, it is well known that
\begin{eqnarray}
\langle (\Delta g_{\ell})^2 \rangle \geq 0 \nonumber
\end{eqnarray}
for all $| \psi \rangle \in {\cal H}$. Then, the operators
$(g_{\ell})^2$ always have diagonal form and eigenvalues that can
be equal only to $1$ and $0$ (Serre 1992). Therefore, the maximum
uncertainty of a local measurement is achieved when the second
term in (4), corresponding to the squared expectation value (3),
is equal to zero. It is clear that this condition requires a
special choice of the state $| \psi \rangle \in {\cal H}$.

Following our previous discussions (Can {\it et al} 2002 (a),
Klyachko and Shumovsky 2002, Klyachko 2002), let us define the
{\it maximum entangled state} in (1) by the condition
\begin{eqnarray}
\forall \ell  \quad \quad \langle g_{\ell} \rangle =0, \quad \quad
g_{\ell} \in Lie \quad G_{\ell} \label{5}
\end{eqnarray}
This means that the perfect entanglement of a composite system
provides the maximum uncertainty of all local measurements
performed over all components. In other words, the maximum
entanglement corresponds to a state, in which all projections of
the spin are equal to zero.

Before we begin to discuss this definition in details, let us note
that the coherent states of photons are widely used for decades in
quantum optics. It is interesting that these states can also be
defined in terms of the dynamic symmetry approach (Perelomov
1986). According to Perelomov's analysis, the coherent states
provide the {\it minimum} uncertainty of local measurements. That
is why the coherent states are usually considered as {\it almost
classical states}.

It is clear that the maximum entangled states defined in terms of
condition (5) represent the very reverse case with respect to the
coherent states. Thus, the perfect entangled states, corresponding
to the {\it maximum} uncertainty of local measurements, should be
considered as the {\it fundamentally quantum states}.

Let us return to the example of $N$-qubit system. Then, each
subspace in (1) is spanned by the two vectors
\begin{eqnarray}
e_{\ell}^{(1)} =|+_{\ell}\rangle, \quad \quad \quad
e_{\ell}^{(2)}=|-_{\ell}\rangle, \nonumber
\end{eqnarray}
where $| \pm_{\ell} \rangle$ denotes the spin-up and spin-down
states of the $\ell$-th spin, respectively. The physical
realization of "spin" variable can be chosen differently. For
example, it can be polarization of photons or state of a two-level
atom. In this local basis, the infinitesimal generators of the
$SU(2)$ group have the following form
\begin{eqnarray}
\begin{array}{ll} \sigma_{\ell}^x = & |+_{\ell} \rangle \langle
-_{\ell}|+|-_{\ell} \rangle \langle +_{\ell}|, \\ \sigma_{\ell}^y
= & -i|+_{\ell} \rangle \langle -_{\ell}|+i|-_{\ell} \rangle
\langle +_{\ell}|, \\ \sigma_{\ell}^z = & |+_{\ell} \rangle
\langle +_{\ell}|-|-_{\ell} \rangle \langle -_{\ell}|, \end{array}
\label{6}
\end{eqnarray}
so that
\begin{eqnarray}
\forall \ell  \quad j=x,y,z \quad \quad (\sigma_{\ell}^j)^2={\bf
1}, \nonumber
\end{eqnarray}
where $\bf 1$ is the unit operator.

Consider the simplest case of only two components ($N=2$). Then,
the Hilbert space (1) is spanned by the four base vectors
\begin{eqnarray}
| \psi_{ik} \rangle =e_1^{(i)} \otimes e_2^{(k)}, \quad \quad
i,k=1,2. \nonumber
\end{eqnarray}
Any state in such a space can be represented as follows
\begin{eqnarray}
| \psi \rangle = \sum_{i,k=1}^2 \psi_{ik} |\psi_{ik}\rangle ,
\label{7}
\end{eqnarray}
where the complex coefficients $\psi_{ik}$ obey the standard
normalization condition
\begin{eqnarray}
|\psi_{11}|^2+|\psi_{12}|^2+|\psi_{21}|^2+|\psi_{22}|^2 =1.
\label{8}
\end{eqnarray}
Employing the definition (5) with the measurements defined by (6)
then gives the following set of six equations
\begin{eqnarray}
\left\{ \begin{array}{l}
Re(\psi_{11}\psi_{21}^*+\psi_{12}\psi_{22}^*)=0 \\
Im(\psi_{11}\psi_{21}^*+\psi_{12}\psi_{22}^*)=0 \\
Re(\psi_{11}\psi_{12}^*+\psi_{22}\psi_{21}^*)=0 \\
Im(\psi_{11}\psi_{12}^*+\psi_{21}\psi_{22}^*)=0 \\
|\psi_{11}|^2+|\psi_{12}|^2-|\psi_{21}|^2-|\psi_{22}|^2=0 \\
|\psi_{11}|^2-|\psi_{12}|^2+|\psi_{21}|^2-|\psi_{22}|^2=0
\end{array} \right. \label{9}
\end{eqnarray}
Thus, the state (7) is characterized by eight real coefficients
(absolute values and phases of $\psi_{ij}$), while the
normalization condition (8) together with conditions (9) give only
seven equations. Since one parameter remains free, there are
infinitely many maximum entangled states in the 2-qubit system.

In general, a state of $N$-qubit system is specified by $2^{N+1}$
real parameters, while conditions (5) together with normalization
condition give only $(3N+1)$ equations. Thus, there are infinitely
many maximum entangled states in an arbitrary $N$-qubit composite
system ($N \geq 2$).

It follows from the normalization condition (8) and Eqs. (9) that
\begin{eqnarray}
\left\{ \begin{array}{l} |\psi_{22}|=|\psi_{11}| \\
|\psi_{21}|=|\psi_{12}| \\ |\psi_{11}|^2+|\psi_{12}|^2=1/2 \\ \cos
\left( \frac{\phi_{11}+\phi_{22}-\phi_{12}-\phi_{21}}{2} \right)=0
\end{array} \right. \label{10}
\end{eqnarray}
where $\phi_{ik} \equiv \arg \psi_{ik}$. Consider some realization
of Eqs. (10). It is easily seen that the choice of either
$|\psi_{11}|=0$ or $|\psi_{12}|=0$ leads to 1the conventional
Einstein-Podolsky-Rosen (EPR) and Bell states
\begin{eqnarray}
\begin{array}{lll}
|\psi_{EPR}\rangle & = & \frac{1}{\sqrt{2}} (|+_1-_2 \rangle \pm
|-_1+_2 \rangle ), \\ |\psi_{Bell}\rangle  & = &
\frac{1}{\sqrt{2}}(|+_1+_2\rangle \pm |-_1-_2 \rangle ),
\end{array}  \label{11}
\end{eqnarray}
respectively. The states (11) form a basis in the four-dimensional
Hilbert space.  Let us stress that  each state in (11) contains
only two base vectors $|\psi_{ij} \rangle$ out of four. The
conditions (10) permit us to construct the maximum entangled
states containing all four base vectors. For example
\begin{eqnarray}
 | \psi \rangle =  \frac{1}{2} (|+_1+_1 \rangle +i|+_1-_2
\rangle +i|-_1+_2 \rangle +|-_1-_2 \rangle   \label{12}
\end{eqnarray}
is the maximum entangled two-qubit state.

Let us stress that, from the mathematical point of view, there is
only one maximum entangled state of the two-qubit system, viz the
EPR state. All other maximum entangled states defined by the
conditions (10) are equivalent to the EPR state to within the
action of the dynamic symmetry group. At the same time, these
states can be different from the physical point of view because
they are realized under different conditions caused by the
physical environment of the system.

We now note that Eqs. (9) can be obtained in a different way. Let
us note that the coefficients $\psi_{ij}$ in (7) form a $(2 \times
2)$ matrix $[ \psi ]$. Then, it is easily seen that the above
equations express the orthogonality conditions for the parallel
rows and columns of this matrix $[\psi]$ and the condition that
different rows and columns have the same norm. The generalization
of this result is discussed in the next section.

Consider now another example of some considerable importance
provided by the three-qubit states. The simplest case is
represented by the Grinberger-Horn-Zeilinger (GHZ) states
\begin{eqnarray}
|\psi_{GHZ} \rangle = \frac{1}{\sqrt{2}} (|+_1+_2+_3 \rangle \pm
|-_1-_2-_3 \rangle ). \label{13}
\end{eqnarray}
The general three-qubit state has the following form
\begin{eqnarray}
| \psi \rangle = \sum_{i,k,m} \psi_{ikm} e_1^{(i)} \otimes
e_2^{(k)} \otimes e_3^{(m)}, \quad \quad i,k.m=1,2, \label{14}
\end{eqnarray}
where $e_{\ell}^{(i)}$ are the same base vectors as above and the
coefficients $\psi_{ikm}$ obey the normalization condition
\begin{eqnarray}
\sum_{i,k,m} |\psi_{ikm}|^2=1. \label{15}
\end{eqnarray}
In this case, the $(2 \times 3)$ matrix $[ \psi ]$ is specified by
eight complex or sixteen real parameter. In turn, the conditions
(5) together with (15) give only ten equations. Thus, there is
infinitely many maximum entangled three-qubit states.

Through the use of definition (5) with Pauli operators (6), we can
get a number of restrictions on the coefficients $\{ \psi_{ikm}
\}$, providing the entanglement in (14). Leaving aside the general
case, we restrict our consideration by the two examples. Consider
first the state
\begin{eqnarray}
| \psi_1 \rangle = \psi_{111}|+_1+_2+_3 \rangle +
\psi_{121}|+_1-_2+_3 \rangle + \psi_{222}|-_1-_2-_3 \rangle .
\label{16}
\end{eqnarray}
Clearly, this is a nonseparable space in (1) and thus it can be
considered as a candidate for entangled state. Then, the use of
the definition (5) gives
\begin{eqnarray}
\left\{ \begin{array}{lll} |\psi_{111}||\psi_{121}| \cos (
\phi_{111} - \phi_{121} ) & = & 0 \\ |\psi_{111}||\psi_{121}| \sin
(\phi_{111} - \phi_{121} ) & = & 0 \\
|\psi_{111}|^2+|\psi_{121}|^2-|\psi_{222}|^2 & = & 0 \\
|\psi_{111}|^2-|\psi_{121}|^2-|\psi_{222}|^2 & = & 0 \end{array}
\right. \nonumber
\end{eqnarray}
where $\phi_{ikm}$ again denotes the phase of the complex
coefficients. It is seen that the only solution of these equations
is
\begin{eqnarray}
|\psi_{111}|=|\psi_{222}|= \frac{1}{\sqrt{2}} , \quad \quad
|\psi_{121}| =0 . \nonumber
\end{eqnarray}
This solution reduces the state (16) to one of the GHZ states (13)
that definitely obey the definition of entanglement in terms of
the maximum uncertainty of local measurements (see Can {\it et al}
2002 (a)). At any $|\psi_{121}| \neq 0$, the nonseparable state
(16) does not manifest entanglement. It should be stressed in this
connection that the nonseparability by itself is not a sufficient
condition of entanglement (Horodecki {\it et al} 1998).

Consider now another, more symmetric realization of the
three-qubit state (14)
\begin{eqnarray}
| \psi_2 \rangle = \psi_{111}|+_1+_2+_3 \rangle +
\psi_{121}|+_1-_2+_3 \rangle + \psi_{212}|-_1+_2-_3 \rangle +
\psi_{222} |-_1-_2-_3 \rangle . \label{17}
\end{eqnarray}
Through the use of definition (5) and normalization (15) we can
obtain the set of ten equations that can be reduced to the
following conditions
\begin{eqnarray}
|\psi_{111}|^2+|\psi_{121}|^2= \frac{1}{2} , \nonumber \\
|\psi_{222}|=|\psi_{111}|, \nonumber \\ |\psi_{212}|=|\psi_{121}|,
\nonumber \\ \phi_{111}-\phi_{121}-\phi_{212}+\phi_{222}= \pm \pi
+2n \pi. \label{18}
\end{eqnarray}
In contrast to the GHZ states (13), the coefficients here do not
have fixed values but lie on a circle of radius $1/2$. Again, the
state (18) is equivalent to the GHZ state (13) to within the
action of the dynamic symmetry group $SU(2) \times SU(2) \times
SU(2)$. The conditions (5) can also be used to construct the basis
of eight three-qubit maximum entangled states.

\section{Maximum entanglement and the matrix of coefficients in (7)}

The maximum entanglement of a nonseparable state is usually
defined in terms of the reduced density matrix. Viz, the reduced
entropy should have the maximum value, same for all components of
the composite system (e.g., see Scully and Zubairy 1997). It is
then a straightforward matter to show that tis condition follows
from the definition of entanglement in terms of local measurements
(5).

Let us now note that the scheme that has been discussed in the
previous section can be reformulated through the use of the
properties of the matrix of coefficients $[\psi]$. It was shown in
previous section that, in the case of two-qubit system, this
matrix obey a certain condition. Consider now the generalization
of this result.

Let the factor spaces in the Hilbert space (1) be spanned by the
orthonormal bases $\{ e_{\ell}^{(i)} \}$. Then, a state of a
composite system defined in (1) can be described by the normalized
state vector of the form
\begin{eqnarray}
| \psi \rangle = \sum \psi_{i_1i_2 \cdots i_N}e^{(i_1)}_1 \otimes
\cdots \otimes e^{(i_N)}_N . \label{19}
\end{eqnarray}
The results of the previous section show us that the entanglement
of the state (19) is specified by a certain choice of the
many-dimensional matrix $[\psi]$ of the coefficients in (19).

It has been proven (Klyachko 2002) that the state $| \psi \rangle
\in {\cal H}$ manifests the maximum entanglement if and only if
parallel slices of its matrix $[ \psi ]$ are orthogonal and have
the same norm. (About parallel slices of multidimensional matrices
see Gelfand {\it et al} 1994. In the simplest case of two-qubit
system considered in the previous section, the parallel slices are
represented by rows and columns of the $(2 \times 2)$ matrix
$[\psi]$. In the case of three-qubit system, this is a $(2 \times
3)$ matrix.)

This general statement can be illustrated in the simplest case by
the state (12) whose matrix of coefficients has the form
\begin{eqnarray}
[\psi]= \left[ \begin{array}{cc} 1/2 & i/2 \\ i/2 & 1/2
\end{array} \right] . \nonumber
\end{eqnarray}
It is clear that
\begin{eqnarray}
[1/2 \quad i/2] \left[ \begin{array}{c} -i/2 \\ 1/2 \end{array}
\right] = 0, \nonumber
\end{eqnarray}
so that the parallel slices are orthogonal. In turn
\begin{eqnarray}
||[1/2 \quad i/2]||=||[i/2 \quad 1/2]||=1/ \sqrt{2}. \nonumber
\end{eqnarray}
The fact that (12) represents the maximum entangled state can also
be verified through the calculation of reduced entropies. It is
straightforward to show that, in the case of state (12)
\begin{eqnarray}
S_1=S_2= \ln 2, \nonumber
\end{eqnarray}
as all one can expect for the maximum entangled two-qubit state
(Scully and Zubairy 1997). Here
\begin{eqnarray}
S_k=-Tr_{i \neq k}(\rho \ln \rho), \quad \quad \rho=|\psi \rangle
\langle \psi| \nonumber
\end{eqnarray}
is the reduced entropy. The above condition of maximum
entanglement together with the Eqs. (10) permits us to construct
another maximum entangled two-qubit states that involve all four
base vectors $\{ e_{ell}^{(1)},e_{\ell}^{(2)} \}$ at $\ell =1,2$.
For example, the states
\begin{eqnarray}
|\psi' \rangle & = & \frac{1}{2} (|+_1+_2 \rangle -i|+_1-_2
\rangle +i|-_1+_2 \rangle -|-_1-_2 \rangle ), \nonumber \\ |
\psi'' \rangle & = & \frac{1}{2} (i|+_1+_2 \rangle +|+_1-_2
\rangle +|-_1+_2 \rangle +i|-_1-_2 \rangle ) , \nonumber \\
|\psi''' \rangle & = & \frac{1}{2} (-i|+_1+_2 \rangle +|+_1-_2
\rangle - |-_1+_2 \rangle +i|-_1-_2 \rangle ) \nonumber
\end{eqnarray}
are the maximum entangled two-qubit states, forming together with
(12)  an orthonormal basis in the Hilbert space (1) (Can {\it et
al} 2002 (a)). This basis is equivalent to (11) to within the
action of the dynamical symmetry group $SU(2) \times SU(2)$.

In a more general case of a two-component entangled state
\begin{eqnarray}
|\psi \rangle \in {\cal H}={\cal H}_1 \otimes {\cal H}_2,
\nonumber
\end{eqnarray}
the matrix of coefficients $[\psi]$ has the dimensionality $n_1
\times n_2$ where $n_{\ell} \equiv \dim {\cal H}_{\ell}$. The
orthogonal rows and columns have the norms $1/\sqrt{n_1}$ and $1/
\sqrt{n_2}$, respectively. Thus, the maximum entanglement is
allowed only if $n_1=n_2$. In this case, $[\psi]$ is similar to
the unitary matrix. This implies the uniqueness of the maximum
entangled state to within the action of the dynamic symmetry group
$SU(n) \times SU(n)$.

By performing a similar analysis, it is a straightforward matter
to show that the three-qubit state (17), (18) also manifests the
maximum entanglement.

Our consideration so far have dealt with the composite systems of
spin-$1/2$ particles (qubit systems). The scheme can be
generalized on the case of an arbitrary spin $s \geq 1/2$ as well.
The examples of the spin-$1$ entangled states were discussed by
Burlakov {\it et al} (Burlakov {\it et al} 1999) in the context of
photon pairs in symmetric Fock states and by Can {\it et al} (Can
{\it et al} 2002 (a)) in connection with polarization of multipole
waves of photons. The application of such a states to the quantum
cryptography was considered by Bechman-Pasquinucci and Peres
(Bechman-Pasquinucci and Peres 2000).

According to the definition of maximum entanglement (5) discussed
in section 2, the entangled state should give the average spin
projection onto every direction equal to zero. Let us denote the
spin-$1$ states by $+ \rangle$, $|0 \rangle$, and $|-\rangle$.
Consider the cascade decay of a two-level atom with the excited
state specified by the angular momentum $j=2$ and projection of
the angular momentum on the quantization axis $m=0$, and the
ground state $j'=0, \quad m'=0$. This transition gives rise to
photon twins (Mandel and Wolf 1995) that can be observed in the
states
\begin{eqnarray}
|+_1-_2\rangle, \quad \quad |0_10_2 \rangle , \quad \quad |-_1+_2
\rangle  \label{20}
\end{eqnarray}
because of the conservation of the total angular momentum in the
process of radiation. It is then easily seen that the so-called
$SU(2)$ phase states of photons (Shumovsky 2000)
\begin{eqnarray}
| \phi_k \rangle = \frac{1}{\sqrt{3}} (|+_1-_2 \rangle +e^{i
\phi_k}|0_10_2 \rangle +e^{2i \phi_k}|-_1+_2 \rangle , \label{21}
\end{eqnarray}
where
\begin{eqnarray}
\phi_k = \frac{2k \pi}{3} , \quad \quad k=0,1,2, \nonumber
\end{eqnarray}
obey the condition (5) and form a basis of entangled states dual
to (20).

The principle difference between the systems with spin $1/2$ and
spin $1$ is that the maximum entangled state (in the sense of
definition (5)) can be realized in the composite system,
consisting of at least two particles in the former case and in a
single-particle system in the latter case. Consider, for example,
the superposition state
\begin{eqnarray}
| \psi \rangle = \lambda_+ |+ \rangle + \lambda_0 |0 \rangle +
\lambda_- |- \rangle, \quad \quad \sum_i|\lambda_i|^2 =1.
\label{22}
\end{eqnarray}
Then, the measurement of projections gives in view of the
definition (5) the following equations
\begin{eqnarray}
\left\{ \begin{array}{l} |\lambda_+|^2-|\lambda_-|^2=0 \\
|\lambda_+||\lambda_0|
\cos (\varphi_0 - \varphi_+)+ |\lambda_0||\lambda_-| \cos (\varphi_- - \varphi_0) =0 \\
|\lambda_+||\lambda_0| \sin (\varphi_0 -
\varphi_+)+|\lambda_0||\lambda_-| \sin (\varphi_- - \varphi_0 )
\end{array} \right. \label{23}
\end{eqnarray}
where $\varphi_i \equiv \arg \lambda_i$. One of the possible
solutions, manifesting the single spin-$1$ particle entanglement
then is $|\lambda_0|=0$ and
\begin{eqnarray}
|\psi_I \rangle = \frac{1}{\sqrt{2}}(|+\rangle +e^{i
\phi}|-\rangle ), \label{24}
\end{eqnarray}
where $\phi$ is an arbitrary complex number. Another solution has
the form $|\lambda_+|=|\lambda_-|=0$, $|\lambda_0|=1$, so that
\begin{eqnarray}
|\psi_{II} \rangle = |0 \rangle . \label{25}
\end{eqnarray}
One more solution of (23) is specified by the conditions
\begin{eqnarray}
|\lambda_+|=|\lambda_-|, \quad \quad
2|\lambda_+|^2+|\lambda_0|^2=1, \quad \quad \cos \left( \frac{
\varphi_++\varphi_--2 \varphi_0}{2} \right)=0. \label{26}
\end{eqnarray}
It is seen that the conditions (26) permit us to construct
infinitely many entangled single-particle states. For example,
\begin{eqnarray}
|\psi_{III} \rangle =|\lambda_+|^2 \left( |+\rangle +
\frac{1+i}{\sqrt{2} |\lambda_+|^2} \sqrt{1-2|\lambda_+|^2} |0
\rangle +|-\rangle \right) \label{27}
\end{eqnarray}
is the single-particle entangled state. From the physical point of
view, these states can be constructed for the  massive particles
like $\rho$ and $K$ mesons that have reasonable long life time and
for the alcaline atoms used in the experiments on Bose-Einstein
condensation. The problem of interpretation, preparing entangled
single-particle states, and performing the necessary measurements
deserves special consideration.

\section{Entangled states in atomic systems}

As a possible physical realization of the above discussed
formalism,  consider now the entangled states in the atomic
systems. It should be stressed that the engineered entanglement in
the systems of trapped atoms and atomic beams has attracted
recently a great deal of interest (e.g., see Bederson and Walther
2000, Mayatt {\it et al} 2000, Rempe 2000, Raymond 2001, Julsgaard
2001 and references therein). In particular, the single-photon
exchange between the two two-level atoms in a cavity can lead to a
maximum entangled atomic state (Plenio {\it et al} 1999, Beige
{\it et al} 2000).

It was then shown (Can {\it et al} 2002 (a)) that the atomic
entangled states in a cavity belong to a special class of the
so-called $SU(2)$ phase states that has been introduced by Vourdas
(Vourdas 1990) and generalized by one of the authors (see
Shumovsky 2001 and references therein). It particular, it was
shown that these states obey the definition of maximum
entanglement (5). The $SU(2)$ phase states can also be used in
quantum coding (Vourdas 2002).

One of the important requirements dictated by the practical
applications of entanglement in the field of quantum information
technologies is that the lifetime of an entangled state should be
long enough.  Under this conditions, it seems to be much more
convenient to use the three-level atoms with the $\Lambda$-type
transitions instead of the two-level atoms (Can {\it et al} 2002
(b)). Consider this idea in more details.

The system of three-level $\Lambda$-type atoms interacting with
the two cavity modes can be described by the following Hamiltonian
\begin{eqnarray}
H=H_0+H_{int}, \label{28} \\ H_0= \omega_P a^+_Pa_P + \omega_{Sk}
a^+_{S}a_{S} + \omega_{12} \sum_f R_{22}(f)+ \omega_{13} \sum_f
R_{33}(f), \nonumber
\\ H_{int}= \sum_k \sum_f  \{g_PR_{21}(f)a_P+g_SR_{32}(f)a_S+H.c. \}
. \nonumber
\end{eqnarray}
Here $a_P$ and $a_{S}$ are the photon operators of the "pumping"
and Stokes modes, respectively. It is supposed that the cavity is
an ideal one with respect to the pumping, while strongly absorbes
the Stokes photons. The operator $R_{21}(f)$ describes the
transition in $f$-th atom from the ground to the highest excited
level. In turn, $R_{32}(f)$ gives the transition from the highest
excited level to an intermediate level $3$ separated from the
ground level by $\omega_{13}$. The dipole transition between the
atomic levels $3$ and $1$ is forbidden because of the parity
conservation.

Assume first that the system consists of only two atoms and is
initially prepared in the state
\begin{eqnarray}
| \Psi_0 \rangle =|1,1 \rangle \otimes |1_P \rangle , \label{29}
\end{eqnarray}
so that both atoms are in the ground state while the cavity
contains a single photon of the pumping mode. The evolution of the
system in the cavity damped with respect to the Stokes photons is
then described by the master equation
\begin{eqnarray}
\dot{\rho}=-i[H,\rho]+ \kappa \{ 2a_S \rho a^+_S -a^+_Sa_S \rho -
\rho a^+_Sa_S \} , \label{30}
\end{eqnarray}
where $1/ \kappa$ is the lifetime of a Stokes photon in the cavity
defining the quality factor. The so-called Liouville term in the
right-hand side of (30) takes into account the absorption of
Stokes photon. The density matrix $\rho$ involves all eigenstates
of the Hamiltonian (28) including the state
\begin{eqnarray}
| \psi_{fin} \rangle = \frac{1}{\sqrt{2}} (|3,1 \rangle +|1,3
\rangle ) \otimes |0_P \rangle \otimes |0_S \rangle . \label{31}
\end{eqnarray}
The lifetime of this state is determined by the nonradiative
processes and therefore is quite long.

As  a consequence of evolution generated by the Hamiltonian (28),
the pumping photon can be absorbed by either atom with equal
probability, so that the system passes into the entangled state
\begin{eqnarray}
| \psi_1 \rangle = \frac{1}{\sqrt{2}} (|2,1 \rangle +|1,2 \rangle
) \otimes |0_P \rangle \otimes |0_S \rangle . \nonumber
\end{eqnarray}
The lifetime of this state is completely defined by the dipole
radiative processes $2 \rightarrow 1$ and $2 \rightarrow 3$ and
therefore is very short.

 As the next step, the first term in the right-hand side of
(30) generates evolution to another entangled state
\begin{eqnarray}
| \psi_2 \rangle = \frac{1}{\sqrt{2}} (|3,1 \rangle +|1,3 \rangle
) \otimes |0_P \rangle \otimes |1_S \rangle . \nonumber
\end{eqnarray}
The absorption of the Stokes photon described by the Liouville
term in (30) then leads to the final, long-lived state (31).

The scheme can be easily realized with the modern experimental
technique. First of all, the single-photon excitation of the
cavity field can be prepared (see Walther 1997, Walther 2001 and
references therein). One of the atoms can be trapped inside the
cavity, while the other atom should pass through the cavity in the
same way as in the experiments on excitation of Fock states of
photons (Walther 2001). Another way is to send a beam of
three-level atoms through the cavity with single pumping photon so
that every time there would be just two atoms inside the cavity.

Concerning the experimental realization, let us note that the
Raman-type process with emission of Stokes photon in a single atom
has been observed recently (Henrich {\it et al} 2000).

In principle, the process can be realized in the system of more
than two three-level atoms, interacting with single cavity photon.
In fact, the Fock states with more than one photon have been
successfully generated (Walther 2001). The use of the definition
of entanglement (5) then shows that if the number of pumping
photons in the cavity is $n$, then the number of atoms,
interacting at once with these photons should be $2n$ (Can {\it et
al} 2002 (a)). In this case, the entangled atomic states can be
constructed as the $SU(2)$ phase states have been discussed by Can
{\it et all} (Can {\it et al} 2002 (a)).

Another realization of long-lived entanglement in the system of
three-level atoms that has been considered by Can {\it et al} (Can
{\it et al} 2002 (b)) assumes that the Stokes photons can leave
the cavity freely. In this case, detection of Stokes photons
outside the cavity can be considered as a signal that the
long-lived atomic entangled state was created.

\section{Conclusion}

Let us briefly discuss the obtained results. The general scheme
has been discussed in sections II and III has the following
structure. To specify the entangled states of a composite system
defined in the Hilbert space  (1) it is necessary: \\ 1) to
specify the dynamic symmetry group $G$ of the factor spaces in
(1);
\\ 2) to specify the local measurements defined by the Lie
algebra, corresponding to the dynamic symmetry group $G$; \\ 3) to
apply the condition (5) that determines the matrix of coefficients
of a general state in (1), corresponding to the entanglement
(maximum entanglement).

The maximum entangled state can also be specified by the condition
that the parallel slices of matrix $[\psi]$ are mutually
orthogonal and have the same norm. It can be proven that this
definition entails the conventional condition expressed in terms
of reduced entropy (Klyachko 2002).

As it follows from the definition (5), the entangled states show
the maximum level of quantum fluctuations in all local
measurements. Therefore, they should be considered as the
fundamentally quantum states in contrast to the almost classical
coherent states, showing the minimum of quantum fluctuations.

The scheme discussed in section II superposes the elements of the
operational approach (zero result for all local measurements
allowed for an entangled state) with the deep mathematics lying
behind the definition of entanglement (5). In particular, it is
possible to show that an arbitrary entangled state (not necessery
the maximum entangled state) can be defined to be the semistable
vector in the Hilbert space $\cal H$ (1) and that the rate of
entanglement can be specified by the length of minimal vector in
complex orbit of entangled state (Klyachko 2000).

It is interesting, that the definition of entanglement represented
by the condition (5) permits us to consider the single-particle
entangled states in the case of spin $s \geq 1$ in addition to the
conventional composite-system states.

The practical realization of long-lived easy monitored
entanglement discussed in section IV seems to be accessible with
the present experimental technique. Let us note that, instead of
the three-levels interacting with the cavity mode, another
environment can be used. An interesting example is provided by the
system of atoms in the presence of dispersive and absorbing
objects (Dung {\it et al} 1998, Dung {\it et al} 2002, Welsch {\it
et al} 2002).

The definition of entanglement in terms of condition (5) is a
general one and may exceed the limits of quantum optics and
quantum information. For example, the combinations of quarks
corresponding to $\pi$ mesons can be treated in terms of the
possible states with the dynamic symmetry provided by the hadron
group $SU(3)$. Then, the definition (5) shows that $\pi^0$
corresponds to the entangled combination of quarks, while
$\pi^{\pm}$ are specified by the coherent combinations (Klyachko
2002). It seems to be tempting to associate the short lifetime of
$\pi^0$ with respect to $\pi^{\pm}$ by the strong quantum
fluctuations in the entangled state and very weak in the coherent
state.

The authors would like to thank Dr. A. Beige, Prof. J.H. Eberly,
Prof. P.L. Knight, Prof. A. Vourdas, Prof. D.-G. Welsch, and Prof.
A. Zeilenger for useful discussions.

\section*{References}

\parindent 0pt

Bechman-Pasquinucci H and Peres A 2000 {\it Phys. Rev. Lett.} {\bf
85} 3313 \\
Bederson B and  Walther H 2000 {\it Advances in Atomic, Molecular,
and Optical Physics}, Vol. {\bf 42} (New York: Academic Press) \\
Beige A, Munro WJ and Knight PL 2000 {\it Phys. Rev. A} {\bf 62}
052102 \\
Bowmeester D, Ekert AK and Zeilinger A 2000 {\it The Physics of
Quantum Information} (Berlin: Springer-Verlag) \\
Brukner \v{C}, \v{Z}ukowski M and Zeilinger A 2001 {\it E-print
quant-ph/0106119} \\
Burlakov AV, Chekhova MV, Karabutova OA, Klyshko DN and Kulik SP
1999 {\it Phys. Rev. A} {\bf 60} R4209 \\
Can MA, Klyachko AA, and Shumovsky AS 2002 a {\it Phys. Rev. A}
{\bf 66}, 022111 \\
2002 b {\it Appl. Phys. Lett.} {\bf 81} 5072 \\
Dung HT, Kn\"{o}ll L and Welsch D-G 1998 {\it Phys. Rev. A} {\bf
57} 3931 \\
Dung HT, Scheel S, Welsch D-G and Kn\"{o}ll L 2002 {\it J. Optics
B} {\bf 4} S169 \\
Gelfand IM, Kapranov MM and Zelevinsky A.V. 1994 {\it
Discriminants, Resultants, and Multi-Dimensional Determinants}
(Boston: Birkhauser) \\
Gisin N, Ribordy G, Tittel W and Zbinden H 2002 {\it Rev. Mod.
Phys.} {\bf 74} 145 \\
 Greenberger DM, Horne M and Zeilinger A
1998 in {\it Bell's Theorem, Quantum Theory, and Conceptions of
the Universe} (Dordreht: Kluwer) \\
Henrich M, Legero T, Kun K and Rempe G 2000 {\it Phys. Rev. Lett.}
{\bf 85} 5872 \\
 Horodecki M, Horodecki P and Horodecki R 1998
{\it Phys. Rev. Lett.} {\bf 80} 5239 \\ Julsgaard B, Kozhekin A
and Polzik E 2001 {\it Nature} {\bf 413} 400 \\
Klyachko AA 2002 {\it E-print quant-ph/0206012} \\
Klyachko AA and Shumovsky AS 2002 {\it E-print quant-ph/0203099}
\\
Mandel L and Wolf E 1995 {\it Optical Coherence and Quantum
Optics} (New York: Cambridge University Press) \\
Mumford D Fogarty J and Kirwan F 1994 {\it Geometric Invariant
Theory} (Berlin: Springer) \\
Myatt CJ, King BE, Turchette QA,
Sackett CA, Kielpinski D, Itano WH, Monroe C and Wineland DJ 2000
{\it Nature} {\bf 403} 269 \\
Perelomov A 1986 {\it Generalized Coherent States and Their
Applications} (Berlin: Springer) \\
Peres A 1998 {\it Physica Scripta} {\bf 76} 52  \\
Plenio MB,
Huelga SF, Beige A and Knight PL 1999 {\it Phys. Rev. A} {\bf 59}
2468 \\
Raimond JM, Brune M and Haroche S 2001 {\it Rev. Mod. Phys.} {\bf
73} 565 \\
Rempe G 2000 {\it Ann. Phys.} (Leipzig) {\bf 9} 843 \\
Scully MO and Zubairy MS  1997 {\it Quantum Optics} (New York:
Cambridge University Press) \\
Serre JP 1992 {\it Lie Algebras and Lie Groups} (New York:
Sringer-Verlag) \\
Shumovsky AS 2001 in {\it Modern Nonlinear Optics} edited by M.W.
Evans (New York: Wiley) \\
Tombesi P and Hirota O 2001 {\it Quantum Communications,
Computing, and Measurements} (New York: Kluwer Academic/Plenum
Publishers) \\
Verdal V and Plenio MB 1998 {\it Phys. Rev. A} {\bf 57} 1619
\\
Vourdas A 1990 {\it Phys. Rev. A} {\bf 41} 1653 \\ 2002 {\it Phys.
Rev. A} {\bf 65} 042321 \\
Walther H 1997 in {\it Quantum Optics and the Spectroscopy of
Solids} edited  by T.
 Hakio\u{g}lu and A.S. Shumovsky (Dordrecht: Kluwer) \\
2001 in {\it Quantum Communications, Computing, and Measurements}
edited by P. Tombesi and O. Hirota (New York: Kluwer
Academic/Plenum Publishers) \\
Welsch D-G, Dung HT and Kn\"{o}ll L 2002 {\it E-print
quant-ph/0205192} \\
Wigner EP 1931 {\it Gruppentheorie und ihre Anwendungen auf die
Quantummechanik der Atomspectren} (Braunschweig: Vieweg) \\ 1939
{\it Ann. Math.} {\bf 40} 149 \\ 1967 {\it Symmetries and
Reflections} (Bloomington: Indiana University Press) \\
Zeilinger
A. 1999 {\it Rev. Mod. Phys.} {\bf 71} S288

\end{document}